\documentclass[aps,pra,superscriptaddress,12pt,tightenlines,nofootinbib]{revtex4}
\usepackage{amsmath,amsthm,graphicx,amssymb,verbatim,listings}
\usepackage[T1]{fontenc}     
\usepackage{lmodern}         
\usepackage[ pdftex, plainpages = false, pdfpagelabels,
                 pdfpagelayout = useoutlines,
                 bookmarks,
                 bookmarksopen = true,
                 bookmarksnumbered = true,
                 breaklinks = true,
                 linktocpage=all,
                 pagebackref=false,
                 colorlinks = true,
                 linkcolor = BrickRed,
                 urlcolor  = blue,
                 citecolor = BrickRed,
                 anchorcolor = green,
                 hyperindex = true,
                 hyperfigures
                 ]{hyperref}
\usepackage[usenames, dvipsnames]{xcolor}
\usepackage{xifthen} 

\newcommand{\arxiv}[2][]{\ifthenelse{\isempty{#1}}{\href{http://arxiv.org/abs/#2}{{\tt arXiv:\allowbreak{}#2}}} {\href{http://arxiv.org/abs/#2}{{\tt arXiv:\allowbreak{}#2 [#1]}}}}

\newcommand{\booktitle}{\textsl}
\newcommand{\hrefdoi}[2]{\href{https://dx.doi.org/#1}{#2}}

\begin{document}

\title{On Feynman's Discussion of Classical Physics Failing at Specific Heat}

\author{Blake C.\ Stacey}
\affiliation{Physics Department, University of Massachusetts Boston}

\date{\today}

\begin{abstract}
I provide the backstory on how a historical error in the
\booktitle{Feynman Lectures on Physics} was corrected.
\end{abstract}

\maketitle

Don Groom's recent account~\cite{Groom:2021} of editing the
transcripts that became the \booktitle{Feynman Lectures on Physics}
prompted me to reflect on how strange it is that I have an
acknowledgment in a more recent edition, while people who did much
more work and who were much more important in the process of making
the books happen have remained anonymous. (The name of at least one
woman in the story, an editor at Addison-Wesley, has been lost to
history~\cite{Sands:2005}.) I am of the generation that
learned from physicists who knew Feynman; my first encounter with the
Feynman-industrial publishing complex was the warts-and-all portrayal
in Gleick's estimable biography~\cite{Gleick:1992}. Consequently, it
is a little odd that I have any involvement to write about. But there
is a (thankfully short) story to be told, and one with a moral or two.

On occasion, somebody voices the idea that in year $N$, physicists
thought they had everything basically figured out, and that all they
had to do was compute more decimal digits. This turns out to be more
myth than fact.

One classic illustration of how the old guys with the beards knew
their understanding of physics was incomplete involves the
\emph{specific heats of gases.}  How much does a gas warm up when a
given amount of energy is poured into it?  The physics of the 1890s
was unable to resolve this problem.  The solution, achieved in the
next century, required quantum mechanics, but the problem was far from
unknown in the years before 1900. And here we come to the
\booktitle{Feynman Lectures on Physics} (1964), volume 1, chapter 40:
\begin{quote}
The first great paper on the dynamical theory of gases was by Maxwell
in 1859.  On the basis of ideas we have been discussing, he was able
accurately to explain a great many known relations, such as Boyle's
law, the diffusion theory, the viscosity of gases, and things we shall
talk about in the next chapter.  He listed all these great successes
in a final summary, and at the end he said, ``Finally, by establishing
a necessary relation between the motions of translation and rotation
(he is talking about the $\frac{1}{2} kT$ theorem) of all particles
not spherical, we proved that a system of such particles could not
possibly satisfy the known relation between the two specific heats.''
He is referring to $\gamma$ (which we shall see later is related to
two ways of measuring specific heat), and he says we know we cannot
get the right answer.

Two years later, in a lecture, he said, ``I have now put before you
what I consider to be the greatest difficulty yet encountered by the
molecular theory.''  These words represent the first discovery that
the laws of classical physics were wrong.  This was the first
indication that there was something fundamentally impossible, because
a rigorously proved theorem did not agree with experiment.  About
1890, Jeans was to talk about this puzzle again.  One often hears it
said that the physicists at the latter part of the nineteenth century
thought they knew all the significant physical laws and that all they
had to do was to calculate more decimal places.  Someone may have said
that once, and others copied it.  But a thorough reading of the
literature of the time shows they were all worrying about something.
\end{quote}
The lesson is right, but either the name or the date is
wrong. ``Jeans'' can surely only be Sir James Jeans, but then ``1890''
cannot be correct. Jeans was born in 1877 and did not go to Cambridge
until 1896.

A better date, in the next century but still in the earliest years of
the old quantum theory, would be 1904, the first publication of his
book \booktitle{The Dynamical Theory of Gases}, which has a lengthy
discussion of how kinetic theory fails to account for gases' specific
heats.  The nub is on p.~173: ``Our theory has, then, led to a result
which is in flagrant opposition to experiment.''  Jeans' own attempt
at a solution was basically to deny that all the degrees of freedom
were in statistical equilibrium~\cite{Jeans:1905}.  Additional
references can be found in Hudson~\cite{Hudson:1989}.

If you want a statement of this issue which really does date to
``about 1890'', see Peter Guthrie Tait's ``On the Foundations of the
Kinetic Theory of Gases'', \booktitle{Transactions of the Royal
  Society of Edinburgh}, 14 May 1886.  Tait writes of Maxwell,
\begin{quote}
  He obtained, in accordance with the so-called \emph{Law of
    Avogadro,} the result that the average energy of translation is
  the same per particle in each system; and he extended this in a
  Corollary to a mixture of any number of different systems.  This
  proposition, if true, is of fundamental importance.  It was extended
  by Maxwell himself to the case of rigid particles of any form, where
  rotations perforce come in.  And it appears in such a case that the
  whole energy is ultimately divided \emph{equally} among the various
  degrees of freedom.  It has since been extended by Boltzmann and
  others to cases in which the individual particles are no longer
  supposed to be rigid, but are regarded as complex systems having
  great numbers of degrees of freedom. [\ldots\!] This, if accepted as
  true, at once raises a formidable objection to the kinetic theory.
  For there can be no doubt that each individual particle of a gas has
  a very great number of degrees of freedom besides the six which it
  would have if it were rigid:—the examination of its spectrum while
  incandescent proves this at once.  But if all these degrees of
  freedom are to share the whole energy (on the average) equally among
  them, the results of theory will no longer be consistent with our
  experimental knowledge of the two specific heats of a gas, and the
  relations between them.
\end{quote}

This passage can be found on p.~135 of the 1900 reprint collection of
Tait's papers.  ``On the Foundations of the Kinetic Theory of Gases''
was actually a five-part series which appeared in the
\booktitle{Transactions} from 1886 to 1892, so a slightly vague dating
makes a kind of sense, and it would be easy to swap Jeans for Tait
while speaking.

Another candidate who fits the dating less well, but should be noted
nevertheless, is Josiah Willard Gibbs, writing at the end of 1901:
\begin{quote}
In the present state of science, it seems hardly possible to frame a
dynamic theory of molecular action which shall embrace the phenomena
of thermodynamics, of radiation, and of the electrical manifestations
which accompany the union of atoms.  Yet any theory is obviously
inadequate which does not take account of all these phenomena.  Even
if we confine our attention to the phenomena distinctively
thermodynamic, we do not escape difficulties in as simple a matter as
the number of degrees of freedom of a diatomic gas.  It is well known
that while theory would assign to the gas six degrees of freedom per
molecule, in our experiments on specific heat we cannot account for
more than five.  Certainly, one is building on an insecure foundation,
who rests his work on hypotheses concerning the constitution of
matter.
\end{quote}
This from the Preface to his \booktitle{Elementary Principles in
  Statistial Mechanics}.

As my friends will testify, I am a complainer who masquerades as a
historian~\cite{Stacey:2016, Stacey:2018, Stacey:2018b,
  Stacey:2019}. So, in 2013, I wrote a letter to the people in charge
of revising the Feynman red books, and the passage was quickly
corrected~\cite{errata}.

What morals can we draw from this little excursion? First, of course,
there is the lesson that Feynman himself was laying out: the
importance of peeling back what Stephen Jay Gould called the
``textbook cardboard'' of a subject~\cite{Gould:1988}. (I am indebted
to Riley Black for teaching me this term, back in the halcyon days of
science blogging.) The real history is inevitably more intricated than
the caricature we receive from books that rush past it to get to
topics on which homework can be assigned. Another lovely example of
this is that Pauli derived the correct quantum-mechanical energy
levels of the hydrogen atom before Schr\"odiger found the
Schr\"odinger equation, by deducing the quantum counterpart of the
Laplace--Runge--Lenz vector~\cite{Pauli:1926, Waerden:1968}. This is
never how we teach undergraduate quantum mechanics, because
second-year undergraduates do not have the background in classical
physics that Pauli did!

A second lesson loops us back around to where we began, with the
production process of the \booktitle{Feynman Lectures}. This erratum
is indicative of the type of glitch that can slip through
proofreading by students and by physicists, precisely because we are
naturally weak on history.

The \booktitle{Feynman Lectures} are widely regarded to be unsuitable
as textbooks, sometimes I think for better reasons than others. The
difficulties that Feynman's own afterword describes sound to me like
the start-up challenges of any new course that does not yet have a
standard text --- challenges that I've had more than one encounter
with. The \booktitle{Lectures} were distributed without the
accompaniment of problems, which degraded their utility. And while a
separate book of exercises was printed much later~\cite{exercises}, I
find that it illustrates, above all, how hard it is to devise problems
that provide a clear gradation of difficulty to guide the student (or
the instructor) into the material of each chapter.

So, it is only right that we dream big, and wonder how we might do it
all differently! (For life is more fragile by the day, and if our
profession cannot deliver on its promises of either utility or beauty,
why should we even stay in it?) But let us anchor that dream in
practicality. The next time we attempt a panoramic survey of modern
physics, we should pay attention to how we might incorporate the
experience of intervening decades, in matters as mundane as
version-tracking, assigning authorship~\cite{Sokol:2019}, and getting
the eyeballs with the right expertise in front of the text.

\end{document}